\documentclass[usenatbib]{mn2e}
\usepackage{natbib}
\usepackage{psfig}
\usepackage{graphicx}
\usepackage{verbatim}
\usepackage{amssymb}
\usepackage{amsmath}
\usepackage{array}
\author[Palliyaguru et al.]{N. T. Palliyaguru$^{\dagger1}$, M. A. McLaughlin$^{1,2,3}$,  E. F. Keane$^{4,5}$, M.~Kramer$^{4,5}$, A. G. Lyne$^4$, \newauthor
 D.~R.~Lorimer$^{1,2}$, R.~N.~Manchester$^6$, F.~Camilo$^{7}$ and I.~H.~Stairs$^8$ \\
\\ $^{\dagger}$ Enquiries to:
npalliya@mix.wvu.edu \\ $^1$ Department of Physics,
West Virginia University, Morgantown, WV 26506, USA. \\ $^2$ Also adjunct at the National Radio Astronomy Observatory, Green Bank, WV 24944,USA. \\ $^3$ Alfred P. Sloan Research Fellow. \\
$^4$ University of Manchester, Jodrell Bank Centre for Astrophysics, Alan Turing Building, Oxford Road, Manchester M13 9PL, UK. \\
 $^5$ Max Planck Institut f\"{u}r Radioastronomie, Auf
dem H\"{u}gel 69, 53121 Bonn, Germany.
\\ $^6$ CSIRO Astronomy and Space Science, Australia Telescope National Facility, P. O. Box 76, Epping  NSW 1710, Australia.\\
$^7$ Columbia Astrophysics Laboratory, Columbia University, 550 W. 120th Street,New York, NY 10027, USA.\\
$^8$ Department of Physics and Astronomy, University of British Columbia,
6224 Agricultural Road, Vancouver, BC V6T 1Z1, Canada.\\
}

\title[Searches for Periodicities and Randomness in Pulse Arrival Times]{Radio Properties of Rotating Radio Transients I: searches for periodicities and randomness in pulse arrival times}
\begin{document}
\title[Searches for Periodicities and Randomness in Pulse Arrival Times]{Radio Properties of Rotating Radio Transients I: searches for periodicities and randomness in pulse arrival times}
\maketitle
\begin{abstract}
We have analysed the long- and short-term time dependence of the pulse arrival times and the pulse detection rates for eight Rotating Radio Transient (RRAT) sources from the Parkes Multi--beam Pulsar Survey (PMPS).
We find significant periodicities in the individual pulse arrival times from six RRATs. These periodicities range from $\sim$30 minutes to 2100 days and from one to 16 independent (i.e. non--harmonically related) periodicities are detected for each RRAT.
In addition, we find that pulse emission is a random (i.e. Poisson) process on short (hour--long) time scales but that most of the objects exhibit longer term (months--years) non--random behaviour. We find that PSRs J1819$-$1458 and J1317$-$5759 emit more doublets (two consecutive pulses) and triplets (three consecutive pulses) than is expected in random pulse distributions. No evidence for such an excess is found for the other RRATs. There are several different models for RRAT emission depending on both extrinsic and intrinsic factors which are consistent with these properties.
\end{abstract}


\section{Introduction}
\label{intro}
Rotating Radio Transients (RRATs) are neutron stars which were discovered only through their isolated pulses \citep{mll+06}. Some, however, have later been detectable through periodicity searches.
The average intervals between detected pulses range from a few minutes to a few hours and 
 pulses have durations between 2 and 30 ms. Thus far, $\sim$50 RRATs have been identified \citep[][]{hrk+08,dcm+09,klk+09,bb10,kkl+11,bbj+11},
 including the original 11 from \citet{mll+06}.
Periods ranging from 0.1 to 8 seconds have been measured for 29 of these sources.  Period derivatives have been measured for 14, allowing inference of spin--down properties such as characteristic ages and surface dipole magnetic fields \citep[][]{mlk+09,lmk09,kkl+11}. 
The periods and magnetic fields of RRATs are larger than those of normal pulsars, but the distributions of other spin--down
properties such as spin--down energy loss rate and characteristic age are similar \citep{mlk+09}.
 Despite this overall trend, the properties of individual RRATs vary
considerably.
Four RRATs, including PSRs J1826$-$1419 and J1913+1330, have spin--down properties consistent with the bulk of the  normal radio pulsar population and
two others, PSRs J1317$-$5759 and J1444$-$6026, have properties similar to normal, older pulsars. Four others, PSRs J1652$-$4406, J1707$-$4417, J1807$-$2557, and J1840$-$1419, lie just above the radio `death--line' \citep[e.g.][]{cr93,zgd07}. However, some have more unusual
spin--down properties. PSRs J0847$-$4316, J1846$-$0257, and J1854$+$0306,  lie in an empty region of $P-\dot{P}$ space between the normal radio pulsars and isolated neutron stars (XINS) and PSR J1819$-$1458 has a high magnetic field of $5\times10^{13}$~G.

 Because of the difficulties in detecting these sporadic objects,
the total Galactic population of RRATs likely outnumbers that of normal radio pulsars \citep{mll+06}, though it is possible that the populations are evolutionarily related \citep{kk+08}. 
Several ideas have been presented about the nature of the emission from these objects.
It could be similar to
that responsible for the `giant pulses' observed from some
pulsars \citep[e.g.][]{kbm+06}.  It could also be that the sporadic emission is related to the fact that these objects are near the radio `death--line' \citep[e.g.][]{cr93,zgd07} and/or are examples of extreme nulling \citep[e.g.][]{rr09}. The phenomenon has also been attributed to
the presence of a circumstellar asteroid belt \citep{li06,cs08} or a radiation belt as seen in planetary magnetospheres \citep{lm07}. 
Or, perhaps, some are transient X-ray magnetars \citep[e.g.][]{wkg+05}.
 Another idea is that their properties lie at the extreme end of the population of normal radio pulsars. 
 Weltevrede et al. (2006) \nocite{wsrw06} show that  PSR B0656$+$14, a nearby middle--aged pulsar which emits pulses with energies many times its mean pulse energy, would be discovered as a RRAT source if it were farther away. 
 RRATs may also be considered as an extreme case of mode changing \citep[see, e.g.,][]{wmj+06} where the on state is less than or about one pulse period.  
Furthermore, \citet{lhk+10} have recently shown that many pulsars exhibit a two--state phenomenon in which varying pulse profile shapes are correlated with variations in spin--down rates and implied changes in magnetospheric particle density. These changes are quasi--periodic, with timescales ranging from one month to many years. 
 It could be that the RRATs are similar two--state systems, in which the profile changes are so dramatic to make them undetectable in the more common state.

Determining the time variability and/or periodicity of the RRAT pulses is therefore an important diagnostic of the RRAT emission mechanism. While the pulse profile, and, in most cases, pulse intensity changes of  \citet{lhk+10} are quasi--periodic,
 the pulse intensity distributions of normal pulsars and giant--pulsing pulsars are believed to be random over time.
On the other hand, nulling pulsars in general show on and off timescales of more than
one consecutive pulse, indicating largely non--random distributions \citep[see, e.g.,][]{rr09}. 
Radio emitting neutron stars often show  transient spin--down phenomena as well. For instance,
 glitches, or sudden increases in the spin frequency, have been observed from young pulsars and one RRAT ( PSR J1819$-$1458). One of the glitches from PSR J1819$-$1458 was accompanied by a 3.5$\sigma$ increase in the pulse detection rate \citep{lmk09}.
Radiative events do not normally accompany the glitches of normal radio pulsars, but are quite common for magnetars \citep{dkg08}.
This, along with the high magnetic field of  PSR J1819$-$1458, hints at a relationship with magnetars and also
 provides additional motivation to examine the pulse rate variations with time for all RRATs. 

We search for periodicities and quantify the randomness of the detected RRAT pulses in several different ways. We first search for periodicities in the pulse arrival times on minutes--year long time scales and pulse detection rates on month--year long time scales using a Lomb--Scargle analysis. We then
quantify the randomness of the RRAT pulse arrival times using Kolmogorov--Smirnov tests on seconds--year long time scales. The
observations are described in Section~2, the methods and results in Section~3, and the conclusions and plans for future work in Section~4.

\section{Observations}
\label{obs}
All eight sources discussed in this paper were discovered by McLaughlin et al. (2006) in a re--analysis of data from the Parkes Multi--beam Pulsar Survey (PMPS). We have ignored three of the original 11 RRATs as their pulse detection rates are too low to perform this analysis.
The discovery data were taken between Jan 1998 and Feb 2002 and
follow--up observations began in  Aug 2003 and are ongoing using the 64-m Parkes telescope. Most of the observations used the central beam of the multi--beam receiver with a central frequency of 1.4 GHz and a bandwidth of 256 MHz. A few observations used other frequencies; we ignore these in our analysis to ensure uniformity of pulse detection rates.
The sources have been observed at between 27 and 89 epochs at 1.4~GHz, with each observation 0.5$-$2~hr in duration (see Table~1).

One important consideration in our analysis is the influence of the interstellar medium on our observed  pulses.
For all of these sources, the predicted diffractive scintillation bandwidths at 1.4~GHz are less than 1~MHz \citep{cl02}, making modulation due to diffractive scintillation unimportant.
In Table~1, we  list the predicted timescales for refractive scintillation at our observing frequency of 1.4~GHz,
estimated from the predicted diffractive scintillation timescales and bandwidths from  \citet{cl02} (see, e.g., \citet{lk05}). These timescales range from 21 to 197 days.
However, the actual timescales could vary significantly from those predicted. The
predicted modulation indices due to refractive scintillation \citep{lk05} range from  0.09 to 0.17, meaning these are expected to be relatively minor contributions to
pulse rate variations.

\section{Analysis and Results}
 
 Pulse detection is performed by dedispersing the data at the dispersion measure (DM) of the RRAT and at a DM of zero. Then
pulses are searched for in both time series above
a 5$\sigma$ threshold using the pulsar processing package
SIGPROC\footnote{http://sigproc.sourceforge.net}. Pulses which are brighter at the DM of the RRAT are likely to be from the source. We inspect the pulses visually by checking for pulse shape and pulse phase consistency
to be certain of their astrophysical nature.
 For some epochs which have large amounts of radio frequency interference, we applied the above procedure but with multiple trial DMs as described by \citet{mlk+09}.
 If more than one pulse is detected within an observation, a second check based on the known period of the source can be made by requiring
that all pulses have arrival times which differ by integer multiples of the period. For the sources with phase--connected timing solutions, we check that the pulse arrival time is consistent with  the solution.
In Table~1, we list the number of epochs for which pulses were detected for all sources. In Table~3 we list the total number of pulses detected within the entire time span of observations.

\subsection{Periodicity search}
\label{LSA}
The Lomb--Scargle test (Scargle et al. 1982) is a statistical procedure for uncovering periodic signals hidden in noise. We use this technique in our analysis as our data are unevenly sampled, thereby making standard Fourier analysis difficult. The implementation we used (Numerical Recipes, see \citet{pftv86}) utilises a version of the periodogram with modifications by Scargle (1982) and Horne and Baliunas (1986). Applications of this test to radio pulsar data can be found in \citet{bjb+97} and \citet{klo+06}. 

The Lomb--Scargle test reveals signals in the power spectral density distribution of a source,
with the presence of a sinusoid of certain frequency indicated by a peak in the spectrum at that particular frequency.
The trial frequencies at which the periodogram is evaluated are chosen to be a finite evenly spaced set. For a time series $X(\rm t_{i})$ with $N_{\rm 0}$ number of elements where $\rm i$ = 1,2,\ldots$N_{\rm 0}$, the Scargle angular frequencies range from $\omega = 2\pi/T$ to $\omega = \pi N_{\rm 0}/T$ (or periods from $T$ to $2T$/$N_{\rm 0}$),  where \textit{T} is the total time interval.  The searched frequencies therefore range up to the Nyquist frequency.
The number of frequencies searched is obtained from the empirical formula (Horne and Baliunas 1986)
\begin{equation}
N_i = -{6.362} + {1.193}\times{N_0} + {0.00098}\times{N_0^2}.
\end{equation}
The likelihood of the existence of a signal or the level of significance is calculated as a detection threshold $Z_{\rm 0}$ (Scargle 1982),
\begin{equation}
Z_0 = -{\rm ln}[1-(1-p_0)^{1/N_i}].
\end{equation}
The false alarm probability  $p_0$ is the probability that a peak of power $Z_{\rm 0}$ will occur in the absence of a periodic signal.

\begin{figure*}\includegraphics[angle=0,width=14cm]{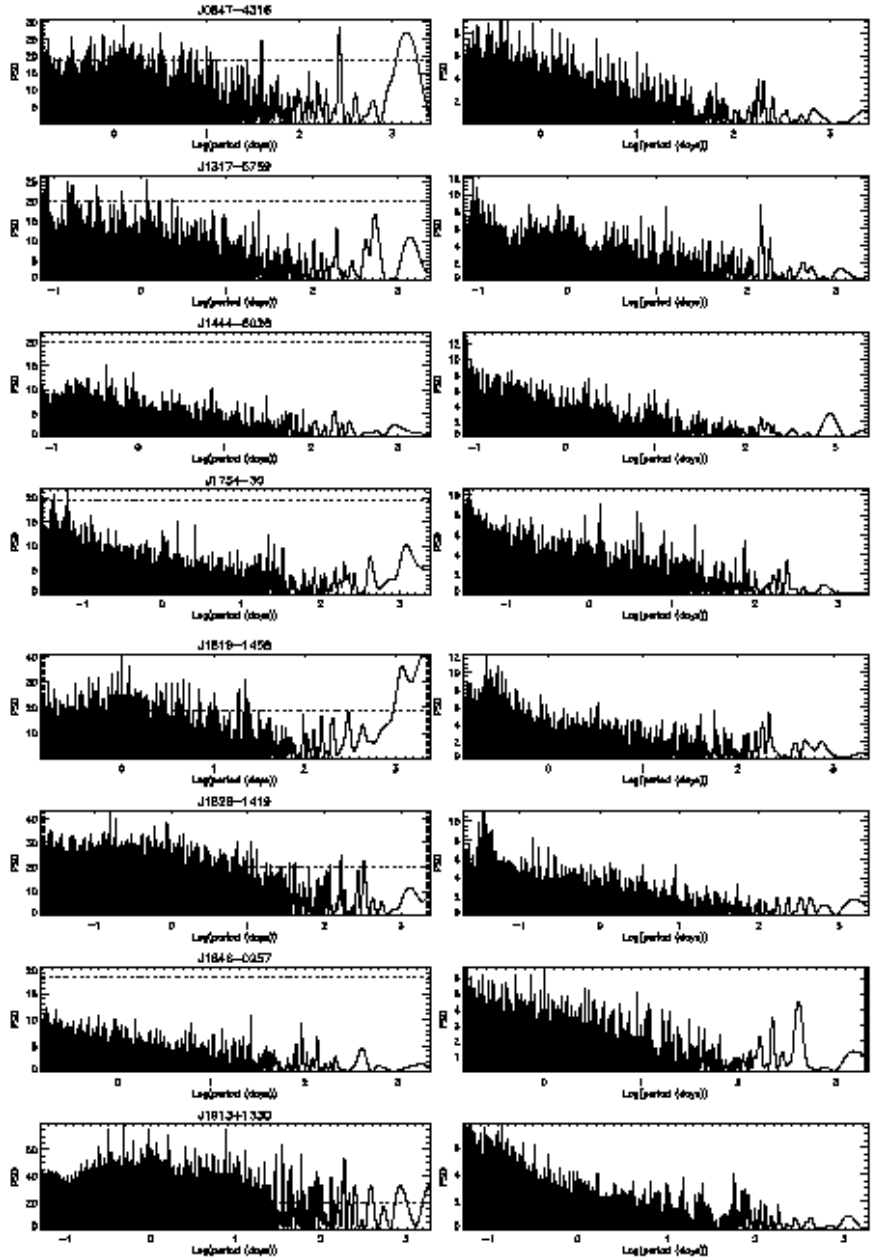}
\caption{Power spectral density vs. period from the Lomb--Scargle analysis on pulse arrival times for PSRs J0847$-$4316, J1317$-$5759, J1444$-$6026, J1754$-$30, J1819$-$1458, J1826$-$1419, J1846$-$0257, and J1913$+$1330 (left panel) and the corresponding plots when the arrival times are randomised (right panel). The randomised time series do not show significant peaks in the spectra. The dashed line represents the 99\% significance level. Note the different y-axis scales in the left and right panels. }
\end{figure*}

\begin{figure*}\includegraphics[angle=0,width=14cm]{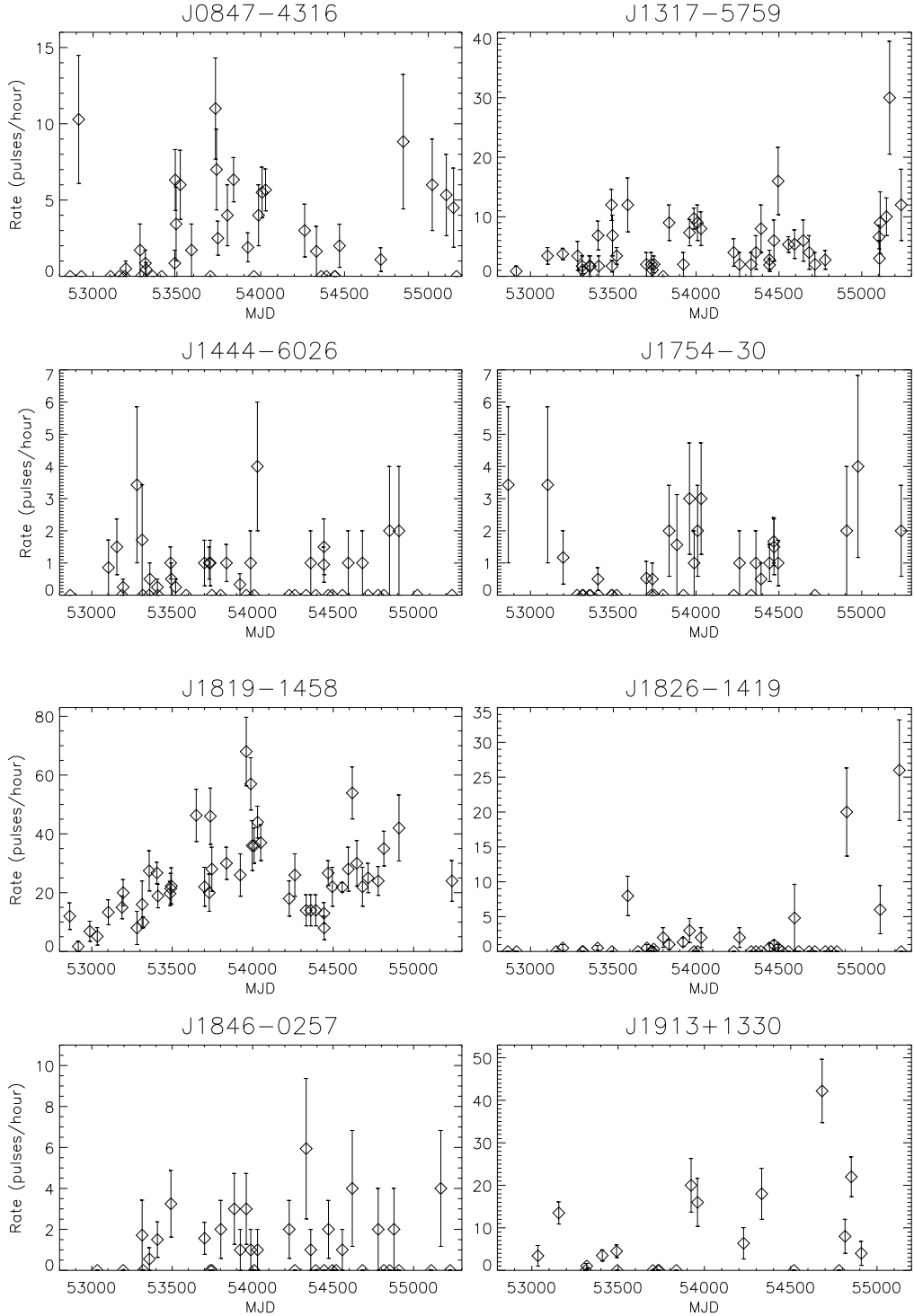}
\caption{Daily pulse detection rate vs. date for PSRs J0847$-$4316, J1317$-$5759, J1444$-$6026, J1754$-$30, J1819$-$1458, J1826$-$1419, J1846$-$0257, and J1913$+$1330.
\label{b1}} \end{figure*}

\begin{figure*}\includegraphics[angle=0,width=14cm]{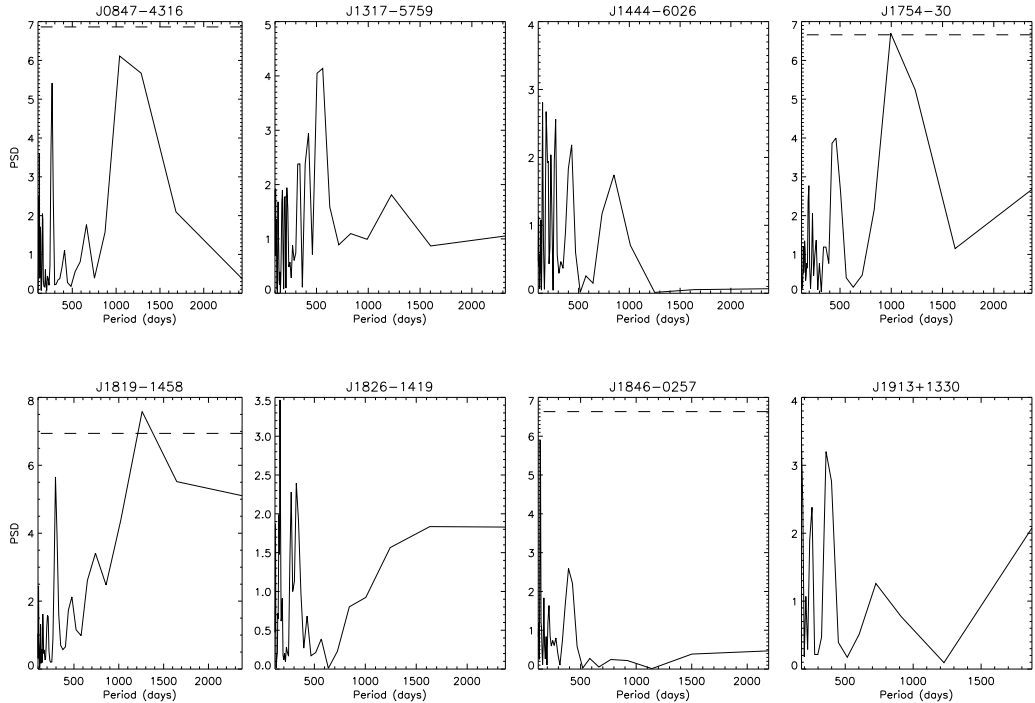}
\caption{Power spectral density vs. period from the Lomb--Scargle analysis on daily pulse detection rates for PSRs J0847$-$4316, J1317$-$5759, J1444$-$6026, J1754$-$30, J1819$-$1458, J1826$-$1419, J1846$-$0257, and J1913$+$1330. The dashed line represents the 95\% significance level.
\label{b2}}
\end{figure*}

\begin{table*}
\label{tb:a1}
\begin{center}\begin{footnotesize}
\caption{Spin period, distance, time between the first and last epochs of observation, the number of epochs, fluctuation period of the pulse detection rate, the significance of the most prominent peak in the spectrum, predicted refractive scintillation timescale (using diffractive scintillation timescales and bandwidths from \citet{cl02}), predicted modulation index due to refractive scintillation, and base-10 logarithms of the derived parameters characteristic age and surface dipole magnetic field strength. }
\begin{tabular}{lllllllllllll}
\hline
Name & Period & Distance & Data Span & N$_{\rm e}$ & $\rm P_{\rm 1}$ & $\rm S_{1}$ & $\Delta$$\rm t_{\rm RISS}$ & $\rm m_{RISS}$ & {$\log[\tau_c]$} & {$\log[B]$} \\PSR & (sec) & (kpc) & (days)  & (days) &(days)& & (days) &  & (yr) & (G)\\
\hline
J0847$-$4316 &5.97& 3.4 & 2438 & 69 & $1040^{+912}_{-167}$ & 89\% & 118 &0.11& 5.9 & 13.4\\
J1317$-$5759 & 2.64& 3.0 & 2324 & 94 & $560^{+68}_{-55}$ & 44\% & 31   &0.15& 6.5 & 12.8\\
J1444$-$6026 &4.76 & 5.5 & 2372 & 72 & 145 $\pm$4 & 5\% & 197 &0.09& 6.6 & 13.0  \\
J1754$-$30 & 1.32& 2.2 & 2372 & 48 & $994^{+239}_{-161}$ & 95\% & 21   &0.17&  -- &  --\\
J1819$-$1458 & 4.26 & 3.6 & 2375 & 72 & $1260^{+386}_{-240}$ & 97\% & 117 &0.11& 5.1 & 13.6\\
J1826$-$1419 & 0.77 & 3.2 & 2375  & 55 & 162$\pm$5 & 24\%  &65  & 0.13& 6.1 & 12.4  \\
J1846$-$0257 & 4.47 & 5.2 & 2193 &  46 & 135$\pm$4 & 89\% & 152 &0.10& 5.6 & 13.4 \\
J1913$+$1330 & 0.92 & 5.7 & 1874 & 27 & 178$\pm$9 & 48\% & 87   &0.13& 6.2 & 12.4  \\
\hline
\end{tabular}
\end{footnotesize}\end{center}\end{table*}

\subsubsection{Periodicity in pulse  arrival times}
 We have searched for periodicities in the pulse arrival times.
In order to do this, we have created a time series by accounting for all rotations of the pulsar during each observation and assigning a delta function (i.e one for each detection and zero for each non-detection). We performed this search on the entire time span of observations. 

 In the time series for the entire time span of observations,
we searched over $\sim$1,000,000 periods spaced as outlined at the beginning of section 3.1, ranging from $\sim$25 minutes to 2300 days.
 Six RRATs PSRs J0847$-$4316, J1317$-$5759, J1754$-$30, J1819$-$1458, J1826$-$1419, and J1913$+$1330, show significant periodicities in the arrival times on these timescales. The peaks of highest significance for PSRs J0847$-$4316, J1317$-$5759, J1754$-$30, J1826$-$1419, and J1913$+$1330 are at 3.8, 1.6, 1.4, 3.6 and 11.3 hours respectively, while PSR J1819$-$1458 shows a long term periodicity of 2102 days. 
There are many other significant periods and harmonics for these RRATs, with
eight, three, one, thirteen, fifteen, and eight independent (i.e. non-harmonically related) periodicities with significance greater than  $99\%$ (2.5$\sigma$) for PSRs J0847$-$4316, J1317$-$5759, J1754$-$30, J1819$-$1458, J1826$-$1419, and J1913$+$1330, respectively (See Table~2). These periodicities range from hours to years.
Because of the large number of detected periodicities, we do not list them all here. 

In order to determine the time dependence of the periodicities,  we divided the time series to halves and quarters and performed the search again.
For every RRAT, all periodicities detected in the full series were re--detected in at least one quarter subsection with lower significances.
None of the periodicities were re--detected in every quarter, though seven out of the eight non--harmonically related periodicities of PSR J1913$+$1330 were re--detected in three of the quarter datasets.
Similarly five out of the 13 non--harmonically related periodicities of PSR J1819$-$1458 were re-detected in three quarters with significances greater than 95$\%$. The rest were re--detected in  only one or two quadrants.

 All detectable (i.e. within the searched range) non--harmonically related periodicities of PSR J1913$+$1330 were re--detected in both halves of the dataset.
 For the rest of the RRATs only about five (for PSR J1819$-$1458) to two (for PSR J1913$+$1330) independent periodicities were re--detected in both halves. 
However every periodicity was re--detected in at least one half section of the dataset with lower significance. These results in general show that the periodicities persist throughout the entire time span of observations.

 In order to gauge the reality of the periodicities, we randomised the time series of detections and non--detections by placing the pulses randomly within the observation windows and repeating the analysis. We found no periodicities with  significance greater than $30\%$ in any of these randomised time series, which suggests the periodicities found are real.
Figure~1 shows the power spectra for the pulse arrival times from the randomised time series for the eight RRATs.


\begin{table*}
\label{tb:adj}
\begin{center}\begin{footnotesize}
\caption{Periodicity in the arrival time $\rm P_{\rm 2}$, and the number of harmonics (nh) detected for each periodicity in the entire time series for the six RRATs. Only the most significant harmonic for each period with a significance larger than $99\%$ is given. The periodicities are listed in the order of their spectral powers. We required the ratio between integer multiples of periods to be less than 1.001 to be considered a harmonic. 
}
\begin{tabular}{|ll|ll|ll|ll|ll|ll|l}
\hline
\multicolumn{2}{|c|}{PSR J0847$-$4316} & \multicolumn{2}{c|}{PSR J1317$-$5759} & \multicolumn{2}{c|}{PSR J1754$-$30} & \multicolumn{2}{c|}{PSR J1819$-$1458} & \multicolumn{2}{c|}{PSR J1826$-$1419} & \multicolumn{2}{c|}{PSR J1913$+$1330}\\
$\rm P_{2}$ & nh& $\rm P_{2}$ & nh& $\rm P_{2}$ & nh& $\rm P_{2}$ & nh& $\rm P_{2}$ & nh&$\rm P_{2}$ & nh \\
(days) & & (days) & & (days) & & (days) & & (days) & &(days) &  \\
\hline
0.163&  10& 0.079& 7&0.066&2&2101.98& 1&0.158&325&0.479&  751\\
1.22& 16&1.16& 1&&&0.996& 33&0.188&      311&7.601&  124\\
265.74& 1&0.138& 6&&&1186.5& 1 &0.869&      143&0.324&  873\\
3.08& 16&0.167& 4&&&1.22&34 &0.026&      969&1.624&437\\
 0.753& 1&&&&&0.907& 38&0.036&      557&0.618&  674\\
 0.343&14&&&&&0.776& 39&0.300&      255&34.657& 28\\
 0.998& 16&&&&&0.548& 36&1.462&      87&0.244&938\\
 0.599&11&&&&&4.668& 26&1.239&      100&1874.01&2\\
&&&&&&0.153& 20&0.048&      490&&\\
&&&&&&3.050& 13&0.407&      213&&\\
&&&&&&19.030& 8&10.934&      14&&\\
&&&&&&6.942& 9& 7.353&      20&&\\
&&&&&&10.202&16&162.892&      1&&\\
&&&&&&&&327.001&      1&&\\
&&&&&&&&117.841&      1&&\\
\hline
\end{tabular}
\end{footnotesize}\end{center}\end{table*}

\subsubsection{Periodicities in daily pulse detection rates}
 We have also applied this method to look for periodicities in the daily pulse detection rates. For each day, the observation length and the number of detected pulses were
used to calculate the rate of pulse detection. Figure~2 shows how this rate varies for the eight RRATs. We then applied the Lomb--Scargle analysis to these
 rates, with the results of this analysis shown in Figure~3. 
We list the most
significant period in Table~1 along with its significance.
 
 We have performed white noise simulations and Monte Carlo simulations to verify the significance of the periodicities. In white noise simulations, the daily rates were replaced by random Gaussian noise. We could then calculate the power spectrum amplitude  corresponding to the desired false alarm probability.
In the second method Monte Carlo simulations were used to generate spectra from random time series which have the same sampling and the cumulative probability distribution of their maximum amplitude is calculated. We then fit this distribution to Equation 2, minimizing $\chi^2$ to determine an effective value for ${\it N_i}$ as this determines the false alarm probability for a given power spectral density. 
These tests verified the significances that we have quoted.

 PSRs J1819$-$1458 and  J1754$-$30 have periodicities with greater than 2$\sigma$ significance at 1260 and 994 days respectively, while PSRs J1846$-$0257 and J0847$-$4316 have periodicities with greater than 1$\sigma$ significance at 135 and 1040 days respectively.
The remaining four RRATs do not show a periodicity of significance greater than the 1$\sigma$ level. The significances obtained for the peak power spectral density for PSRs J1819$-$1458 and J1754$-$30 from white noise simulations are 97\% and 96\% respectively. The significances calculated from Monte Carlo simulations are 97\% and 89\%.

 These periodicities are different from the ones detected in their pulse arrival times except for the $1260^{+386}_{-240}$ day periodicity of  PSR J1819$-$1458, which is close to the pulse arrival time periodicity of 1186$\pm$7 days.
In general the pulse arrival time and  pulse rate recurrence periodicities are expected to be independent though it is possible for them to be the same.
 In order to further gauge the reliability of our results for PSRs J1819$-$1458 and J1754$-$30, we created 1000 random sequences by assigning the measured  pulse detection rates to randomly selected MJDs and calculated the number of times a peak of the same or higher significance appeared in the 1000 trials.  This number was  six and ten  for PSRs J1819$-$1458 and J1754$-$30, respectively, suggesting that the periodicities detected are real and that their significances may in fact be
 underestimated. However, the detected periodicities are at 40\% and 50\% of the total observations lengths for PSRs  J1754$-$30 and J1819$-$1458, respectively (see Table~1); longer observation spans are necessary to determine whether they are real. We also note that the significance of the peaks depends on the ranges of frequencies searched and that we have not searched frequencies higher than the Nyquist frequency or lower than $1/T$.

 In order to determine whether there was any dependence of significance on period, we simulated sinusoidal signals of various periods with additive random Gaussian noise and then applied
 the Lomb--Scargle algorithm. We found that the significance
 of the detected periodicities is independent of period.

\subsection{Randomness tests on pulse arrival times}
\label{random}

The Kolmogorov--Smirnov (KS) test (see, e.g., Press et al. 1986\nocite{pftv86}) is a statistical procedure which determines the degree to which two datasets differ.
 It compares the cumulative probability distributions of both datasets by calculating the maximum deviation
\begin{equation}
D = \max_{-\infty < x < \infty} \mid C_1(x) - C_2(x) \mid,
\end{equation}
of cumulative distribution functions (CDFs) $C_{1}$  and $C_{2}$.
The accuracy of this test increases with the number of data points and is expected to be accurate for
 four or more points \citep{stephens70}. From this the probability that two arrays of data values are drawn from the same distribution can be calculated. Small values of this probability (i.e. large values of $D$)
 suggest that the distributions being tested differ.

The test can therefore be used to explore whether the observed pulse sequences are consistent with random (i.e. Poisson) distributions.
We have done this both by comparing the CDF of our data with the CDF of a simulated randomly distributed pulse sequence (i.e. kstwo test as implemented in Numerical Recipes) and by comparing the CDF of our data directly with the uniform CDF (i.e. ksone test). We have performed this test both for single days and for the entire
time span of observations.
For each single data set, we created 
simulated distributions by placing the same number of detected pulses at random times within the observation, with the constraint of allowing only one detected
pulse per rotation.
We also tested randomness on longer time spans by creating random time sequences of length equal to the total observation lengths.

In Table~3, we list the total number of pulses observed from each RRAT along with the  numbers of pulses detected on individual days. We also list the probabilities of the pulse sequences being random on a single day, averaged over all days of observation by comparing with simulated random pulse sequences and directly with a uniform CDF. The simulated probabilities on each day are averages of 100,000 trials of different randomly generated data sets. 
We also  list the rms deviations of the averages. Table~3 also lists the probability of pulses being randomly distributed on long data spans for simulated data (again averaged over 100,000 realisations)
and through comparison with a uniform CDF. Figure~4 illustrates this comparison for two RRATs.

\begin{table*}
\label{tb:a2}
\begin{center}\begin{footnotesize}
\caption{Total observation time, total number of detected pulses, minimum, maximum, mean (with standard deviation in parentheses) of the number of pulses observed per day, average probability of short term randomness from the simulations (P$_{\rm r1}$), the rms of the average probability ($\sigma$$_{1}$), average probability of long term randomness from the simulations (P$_{\rm r2}$), the rms of the average ($\sigma$$_{2}$), average probability of short term randomness from direct comparison with the uniform CDF (P$_{\rm u1}$), and the average probability of long term randomness from direct comparison with the uniform CDF (P$_{\rm u2}$). The probabilities for the simulations were calculated for 100 trials. The duration of each observation was 30 minutes to 1 hour on average; as can be seen from $\rm N_{2}$ there was one very long observation for PSR J1819$-$1458.
 The days with less than three pulses detected have not been included in P$_{\rm r1}$ or the P$_{\rm u1}$ probability calculation. The numbers outside and inside parentheses for P$_{\rm r1}$, $\sigma$$_{1}$ and P$_{\rm u1}$ indicate those for days with more than two/three pulses detected.}
\begin{tabular}{llllllllllllllll}
\hline
Name & T & N$_{\rm p}$ & N$_{1}$ & N$_{2}$ & N$_{\rm mean}$ & P$_{\rm r1}$ & $\sigma$$_{1}$ & P$_{\rm r2}$  & $\sigma$$_{2}$ & P$_{\rm u1}$ & P$_{\rm u2}$\\PSR & (hr)\\
\hline
J0847$-$4316 & 49 & 141 &0 & 11 & 2.04(2.6) & 0.52(0.53) & 0.15(0.13) & $5\times10^{-3}$ & 0.02  & 0.37(0.37) & $5\times10^{-8}$ \\
J1317$-$5759 & 50 & 256 & 0 &12 & 2.72(2.8) &0.50(0.51) & 0.16(0.15) & $3\times10^{-3}$ & 0.01 & 0.42(0.46) & $1\times10^{-3}$ \\
J1444$-$6026 & 85 & 41 & 0 & 4 & 0.57(1.0) & 0.47(0.45) & 0.11(0.11) & 0.54  & 0.29 & 0.32(0.23) & 0.31 \\
J1754$-$30  & 60 & 40 &  0 & 3 & 0.83(0.9) & 0.09(--) & $2\times10^{-3}$(--) & 0.03 & 0.07 & $3\times10^{-3}$ (--) & $1\times10^{-3}$\\
J1819$-$1458 & 45 & 1102 & 1 & 165 & 15.31(19.5) & 0.49(0.49) & 0.15(0.15) & $3\times10^{-5}$ & $2\times10^{-4}$ & 0.42(0.47) & $2\times10^{-12}$\\
J1826$-$1419 & 53 & 60 & 0 & 13 & 1.09(2.5) & 0.17(0.16) & 0.06(0.05) & $1\times10^{-4}$ & $4\times10^{-4}$ & 0.05(0.04) & $2\times10^{-3}$ \\
J1846$-$0257 & 37 &39& 0 & 4 & 0.85(1.2)  & 0.41(0.29) & 0.17(0.03)  & 0.49 & 0.26 & 0.24(0.06) & 0.46 \\
J1913$+$1330 & 19 &138& 0 & 27 & 5.11(7.8) & 0.18(0.17) & 0.13(0.14) & $6\times10^{-7}$ & $8\times10^{-6}$  & 0.06(0.07) & $2\times10^{-18}$ \\
\hline
\end{tabular}
\end{footnotesize}\end{center}\end{table*}

\begin{figure*}\includegraphics[angle=0,width=14cm]{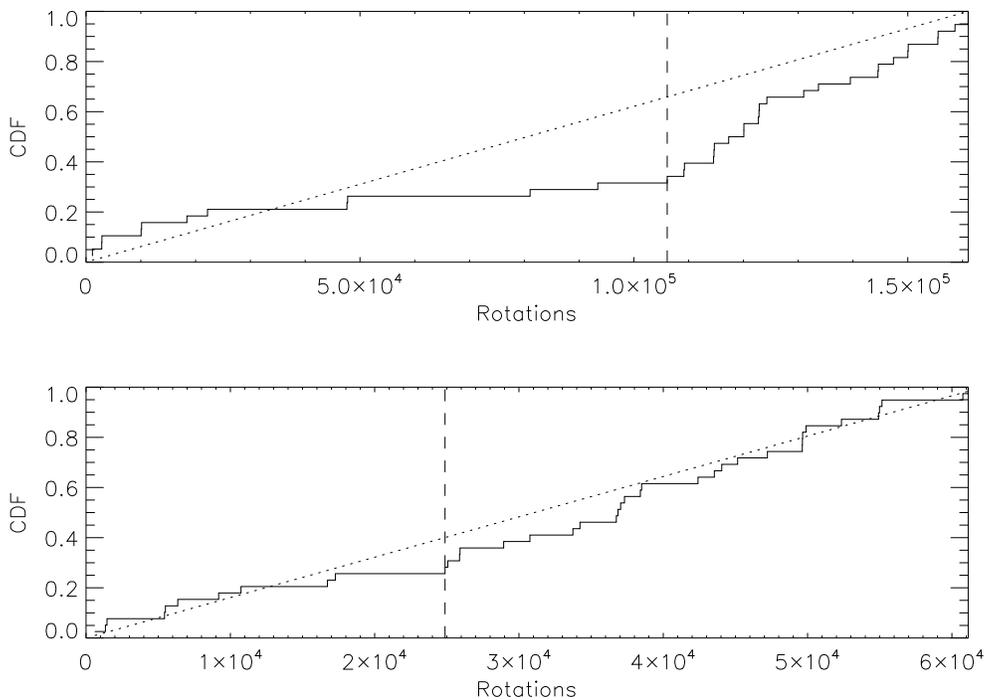}
\caption{Cumulative distribution function for data over the entire time span of observations of PSR J1754$-$30 (an example of a non--random pulse sequence) and PSR J1444$-$6026 (an example of a random pulse sequence) compared with the CDF of a uniform distribution (dashed line). The figure also shows the point at which the KS statistic $D$ is measured (vertical dashed--dotted line). Small values of $D$ imply that the distributions are similar to each other. It can be seen that the first plot differs significantly from a uniform distribution, whereas the second does not. The probabilities of randomness are 0.001 and 0.31 for PSRs J1754$-$30 and J1444$-$6026 respectively.
\label{b3}} \end{figure*}

 In general, the pulse distributions appear to be consistent with random (i.e Poisson) distributions.
 For simulated data, the single--day pulse distributions have probabilities ranging from 0.09 (for PSR J1754$-$30) to 0.52 (for PSR J0847$-$4316) of being random when only days with more than two detections are considered.
  When days with more than three are considered the probabilities range from 0.16 (for PSR J1826$-$1419) to 0.53 (for PSR J0847$-$4316), indicating that the bias for small numbers of pulses is small (there are no epochs with three or more pulses for PSR J1754$-$30).
 When we compare directly with a uniform CDF,  PSRs J1826$-$1419 and J1913$+$1330 show more evidence (i.e. probabilities of 0.05 and 0.06) for non--random pulse emission. 
  PSR J1754$-$30 shows evidence for non--random behavior in both tests. As this is the RRAT with the lowest overall pulse rate, however,
the tests are not very accurate.

 The total number of detected pulses from each source ranges from $\sim$40 to $\sim$1100, making our tests for randomness on long time scales very sensitive.
When long time spans are considered, only  PSRs J1444$-$6026 and J1846$-$0257 are consistent with random distributions. The probabilities given by using the ksone and kstwo tests are similar. This is not
surprising given the large numbers of pulses.
 We have tested these analysis techniques on intervals between pulses and checked for their consistency with an exponential distribution.
Both pulse arrival time and pulse interval tests give identical results as expected. Furthermore we have searched for any correlation between the pulse intervals on different timescales and have found none.

 We explored the possibility of the telescope zenith angle at the time of the observation being responsible for variations in pulse detection rates, as the system temperature depends on this factor and to account for effects based on varying gain and spillover. The 
KS test was carried out on pulse sequences for lower and higher angles (i.e.
angles less and greater than 45 degrees) for this purpose. These results indicate that there is little
dependence of the detected rates on the zenith angle, with probabilities ranging from 0.31 (for PSR J1317$-$5759) to 0.49 (for PSR J1819$-$1458).

We have also searched for clustering of pulses. For all of the RRATs, the pulses usually occur singly with  occasional consecutive pulses detected.
We have measured the total number of doublets, triplets and quadruplets (i.e. instances of two, three, and four consecutive pulses) and compared with the number of doublets, triplets and quadruplets found in simulated random distributions. Table~4 lists the results of our analysis for 1000 simulated distributions.
The number of doublets detected is higher than the number expected for a random distribution for both  PSRs J1819$-$1458 and J1317$-$5759. PSR J1819$-$1458 shows many more doublets, triplets and quadruplets than expected, with one 
instance of nine consecutive pulses.  The mean duration of the  on states (at which pulses are detected) for the eight RRATs in units of the period are also listed in Table~4. These were calculated by taking into account the number of detected multiplets and their duration. For the RRATs which do not exhibit multiplets, we can only give an upper limit to the on--state
duration of one period  as the mean on--state will be less than one period.


\begin{table*}
\label{tb:adj}
\begin{center}\begin{footnotesize}
\caption{Measured number $\rm N_{\rm m}$ of doublets, triplets, quadruplets vs. the expected number $\rm N_{\rm e}$, probability of occurrences under an assumed Poisson distribution $\rm P_{\rm p}$ for eight RRATs and the mean duration of the on state in units of the period. The expected numbers are the means of 1000 trials of randomly generated fake distributions. }

\begin{tabular}{lllllllllll}
\hline

Name & \multicolumn{3}{c}{Doublets} &
\multicolumn{3}{c}{Triplets} & \multicolumn{3}{c}{Quadruplets} & \multicolumn{1}{c}{On state duration}\\
PSR& {$\rm N_{\rm e}$}&
{$\rm N_{\rm m}$} & {$\rm P_{\rm p}$}&{$\rm N_{\rm e}$} &
{$\rm N_{\rm m}$} & {$\rm P_{\rm p}$} & {$\rm N_{\rm e}$} & {$\rm N_{\rm m}$} &{$\rm P_{\rm p}$} & {(upper limit)}\\
\hline
J0847$-$4316 & 1.06  & 1 & 0.35 & 0.004 & 0 & 0.13 & 0 & 0 & 0.06 & 1.007\\
J1317$-$5759 & 1.44  & 7 & 0.003 & 0.009 & 1 & 0.161 & 0 & 0 & 0.021 & 1.036\\
J1444$-$6026 & 0.05 & 0 & 0.949 & 0 & 0 & 0.924 & 0 & 0 & 0.901 & 1\\
J1754$-$30 & 0.014 & 0 & 0.981 & 0 & 0 & 0.971 & 0 & 0 & 0.961 & 1\\
J1819$-$1458 & 35.42  & 111 & $2.5\times10^{-8}$ & 1.3 & 15 & $9.9\times10^{-25}$ & 0.04 & 7 & $3.8\times10^{-45}$ &1.172\\
J1826$-$1419 & 0.06 & 1 & 0.028 & 0.001 & 0 & 0.957 & 0 & 0 & 0.943 & 1.016\\
J1846$-$0257 & 0.06 & 0 & 0.902& 0 & 0 & 0.858 & 0 &0 & 0.815 & 1\\
J1913$+$1330 & 0.97  & 1 & 0.306 & 0.015 & 1 & 0.356 & 0 & 0 & 0.358 & 1.022 \\
\hline
\end{tabular}
\end{footnotesize}\end{center}\end{table*}

\section{Discussion}

 We have shown that  six of the RRATs have periodicities of significance greater than 99$\%$ in the pulse arrival times. The periods of the most significant peak range from 1.4 hours (for PSR J1754$-$30) to 2102 days (for PSR J1819$-$1458). No significant periodicities were detected upon randomizing the time series, showing that these periodicities are real.
We do not find any relationship between the number and significance of detected periodicities and spin--down properties such as period or characteristic age. It is possible that some of the periodic  behavior is due to refractive scintillation. However, the number and wide range of timescales of the periodicities are impossible to explain with refractive scintillation alone.

 The shorter timescale periodicities in pulse arrival times are similar to typically observed nulling timescales, which range from minutes to days \citep{wmj+06}. Explanations for pulsar nulling include an empty sight--line passing through the
sub--beam structure \citep{dr01}, a reversal of the emission direction \citep{mg+06}, pulsar emission ceasing temporarily due to intermittent failure of pair production (Zhang et al. 2007), an asteroid belt of material \citep{cs08} or changes in magnetospheric currents \citep{lhk+10,wmj+06}. Any combination of these could explain the extreme pulse-to-pulse variability of the RRATs and also the longer term periodicities.

 
The significant periodicities found in the RRAT pulse arrival times may suggest a relationship with
pulsars whose spin--down rates and pulse shapes un-
dergo periodic variations, with implied changes in magnetospheric particle density \citep{lhk+10}.
The prototype of this source class is B1931+24 (Kramer et al. 2006). The asteroid belt model of \citep{cs08} attributes the 40-day on/off timescale of
this pulsar to an asteroid with eccentric 40-day orbit.
It may be that the pulse--to--pulse variability of the RRATs is due to a
similar process happening on very short timescales. However, because
the on states of the RRATs are so short, it is impossible to measure
period derivatives during the on and off states. It could also be that
a similar process is causing the longer term periodicities in pulse
arrival times.

 If we apply the model of \citep{cs08} to the RRATs,  multiple asteroids
of an asteroid belt could be responsible for the observed periodicities in the arrival times.
This is consistent with the large root mean square timing residuals which range from 1.1 ms (for PSR J1913$+$1330) to 11.2 ms (for PSR J0847$-$4316) of these RRATs as an earth--sized asteroid would induce residuals of the order of 1 ms \citep{cs08}.
However it is possible that much of these large residuals are due to pulse--to--pulse jitter, indicating that the true asteroid
 mass cannot be determined by the residuals only.

 The periodic fluctuations in pulse arrival times could also be due to non--radial oscillations which drive different emission modes, as often seen in white dwarf stars \citep{rr11}.
The fundamental oscillation periods for neutron stars are expected to be on the order of milliseconds \citep{rg92} to seconds \citep{mh88} for g--modes. 
These are far too short to explain the multiple periodicities seen, but it is possible that we are observing the beat frequency between a non--radial oscillation period and that from a longer timescale process like those observed in \citet{lhk+10}. 

We also searched for periodicities in the
daily pulse detection rates.
Six of the RRATs do not show any significant periodicities in their daily pulse detection rates over timescales of months to years.
 The exceptions are PSR J1819$-$1458, which exhibits a 1260 day period with a significance of 97\%, and PSR J1754$-$30, which exibits a 994 day period with significance of 95\%.  We detect a peak of higher significance only 0.6\% and  1.0\% of the time in 1000 trials in which the rates were randomly assigned to the MJDs for PSRs J1819$-$1458 and J1754$-$30 respectively, suggesting that the significance of the peak may be underestimated by the Lomb--Scargle algorithm.
    Given eight trials, if all of the RRATs had no periodicities, we would expect to find one periodicity with significance greater than 88\%. However, two RRATs with significances greater than 95\% are not expected.
The predicted timescales for refractive interstellar scintillation of 117 days for PSR J1819$-$1458  and 21 days for PSR J1754$-$30 are much smaller than the reported periodicities in pulse detection rates, indicating that these periodicities are not likely due to scintillation.
However, these periodicities are roughly half of the total data span.  Therefore, a longer time span of
 observations is necessary to confirm them as significant.
 
 There is evidence for changes in period derivative associated with changes in pulse detection rate for
PSR J1819$-$1458 \citep{lmk09}. The
 peak in the rate immediately following the first (and largest) glitch at MJD 53926 \citep{lmk09} hints at a correlation
between glitches and emission properties. If confirmed in future glitches,
this correlation might suggest a link with the magnetars, for which radiative changes often accompany glitches \citep[e.g.][]{dkg+08} or the class of mode changing pulsars whose period derivative undergoes quasi--periodic changes \citep{lhk+10}.

All of the sources exhibit random pulse distributions on single days. 
 This is similar to that observed for giant pulses \citep{kbm+06}. However, the pulses from the RRATs are wider than those of observed giant pulses and the pulse amplitude distributions are different \citep{mlk+09,mmk+10}. This therefore supports the idea of Weltrevrede et al. (2006), who suggest that the RRATs could be normal, but more distant, pulsars like
B0656+14 which emits very bright, narrow pulses and a distribution of weaker, broader pulses.

Over longer time spans, the
 pulse sequences  of six of the RRATs show evidence for non--random behaviour. These same RRATs show significant periodicities in arrival times. The external influence from, e.g., an asteroid belt \citep{cs08} could explain
this through uneven distributions of material.
However,
 it is difficult to relate this long term non--randomness to either normal or nulling pulsars as, to our knowledge, this has not been explored for either of
these source classes.
We do not find any obvious relationship between the randomness and derived spin--down properties such as age, period, and surface dipole magnetic field.

Nulling pulsars tend to have on--off states which persist over more than one pulse period \citep{wmj+06} whereas the large majority of the RRAT pulses occur singly, aside from a few pulses from PSRs J1819$-$1458 and J1317$-$5759.
Redman \& Rankin (2009) showed that the majority of pulsars null non--randomly and several studies (e.g.
Rankin \& Wright 2008 and Herfindel \& Rankin 2009) have revealed periodicities in the null cycles for some pulsars.
 The RRATs have quite different properties in these respects, but because the nulls of some pulsars
(e.g.
  B0834$+$06, B1612$+$07, and B2315$+$21; \citet{rr09}) are random and single--pulse nulls are occasionally observed,
 we cannot exclude the possibility of the RRATs being an extreme case of nulling.
 This is supported by the similar ages and spin periods of many RRATs and nulling pulsars and the strong correlation between nulling and pulsar age
\citep{wmj+06}.

 In summary, we detect highly significant periodicities in the arrival times of six of the RRATs. 
We detect no highly significant periodicities in the long--term pulse detection rates for the RRATs, although we find tentative periodicities for PSRs J1819$-$1458 and J1754$-$30. All of the RRATs show random behavior on a single day and
most of the RRATs
show non--random behavior on long timescales. Most of the RRATs emit pulses singly, but a few do show
evidence for clustering of pulses. 
It is clear that there are
periodicities in the pulse arrival times for these objects. The cause
of these could be circumstellar material, non--radial oscillations, or another process.
Radio monitoring over longer time spans and observations at other wavelengths may be useful to further understand the reasons for the RRATs' unusual emission. More theoretical work and further studies of the randomness of pulse emission in normal pulsars with different properties is also necessary.

\section*{Acknowledgments}

We thank the anonymous referee for very useful suggestions.
 We thank all on the Parkes Multi--beam Survey team for assistance with the Parkes radio observations.
MAM, NTP, and DRL are supported by a WV EPSCoR grant.
Pulsar research at UBC is funded by an NSERC Discovery Grant.
EK is supported by the EU Framework 6 Marie Curie Early Stage Training programme under contract number MEST-CT-2005-19669 ``ESTRELA''.

\bibliography{journals,psrrefs2,modrefs2}

\begin{thebibliography}{}

\bibitem[\protect\citeauthoryear{Bailes, Johnston, Bell, Lorimer, Stappers,
  Manchester, Lyne, D'Amico \& Gaensler}{Bailes et~al.}{1997}]{bjb+97}
Bailes M.,  Johnston S.,  Bell J.~F.,  Lorimer D.~R.,  Stappers B.~W.,
  Manchester R.~N.,  Lyne A.~G.,  D'Amico N.,    Gaensler B.~M.,  1997,
  Astrophys.\ J., 481, 386

\bibitem[Burke-Spolaor et al.(2011)]{bbj+11} Burke-Spolaor, 
S., et al.\ 2011, arXiv:1102.4111 

\bibitem[\protect\citeauthoryear{{Burke-Spolaor} \& {Bailes}}{{Burke-Spolaor}
  \& {Bailes}}{2010}]{bb10}
{Burke-Spolaor} S.,  {Bailes} M.,  2010, MNRAS, 402, 855

\bibitem[\protect\citeauthoryear{Chen \& Ruderman}{Chen \&
  Ruderman}{1993}]{cr93}
Chen K.,  Ruderman M.,  1993, Astrophys.\ J., 408, 179

\bibitem[\protect\citeauthoryear{{Cordes} \& {Lazio}}{{Cordes} \&
  {Lazio}}{2002}]{cl02}
{Cordes} J.~M.,  {Lazio} T.~J.~W.,  2002

\bibitem[\protect\citeauthoryear{{Cordes} \& {Shannon}}{{Cordes} \&
  {Shannon}}{2008}]{cs08}
{Cordes} J.~M.,  {Shannon} R.~M.,  2008, ApJ, 682, 1152

\bibitem[Deneva et al.(2009)]{dcm+09} Deneva, J.~S., et al.\ 
2009, ApJ, 703, 2259 

\bibitem[\protect\citeauthoryear{Deshpande \& Rankin}{Deshpande \&
  Rankin}{2001}]{dr01}
Deshpande A.~A.,  Rankin J.~M.,  2001, Mon.\ Not.\ R.\ Astron.\ Soc., 322, 438

\bibitem[\protect\citeauthoryear{{Dib}, {Kaspi} \& {Gavriil}}{{Dib}
  et~al.}{2008a}]{dkg08}
{Dib} R.,  {Kaspi} V.~M.,    {Gavriil} F.~P.,  2008a, in {C.~Bassa, Z.~Wang,
  A.~Cumming, \& V.~M.~Kaspi} ed., 40 Years of Pulsars: Millisecond Pulsars,
  Magnetars and More Vol.~983 of American Institute of Physics Conference
  Series, {10 Years of RXTE Monitoring of Five Anomalous X-Ray Pulsars}.
pp 262--264

\bibitem[\protect\citeauthoryear{{Dib}, {Kaspi} \& {Gavriil}}{{Dib}
  et~al.}{2008b}]{dkg+08}
{Dib} R.,  {Kaspi} V.~M.,    {Gavriil} F.~P.,  2008b, ApJ, 673, 1044

\bibitem[\protect\citeauthoryear{{Hessels}, {Ransom}, {Kaspi}, {Roberts},
  {Champion} \& {Stappers}}{{Hessels} et~al.}{2008}]{hrk+08}
{Hessels} J.~W.~T.,  {Ransom} S.~M.,  {Kaspi} V.~M.,  {Roberts} M.~S.~E.,
  {Champion} D.~J.,    {Stappers} B.~W.,  2008, in {Bassa} C.,  {Wang} Z.,
  {Cumming} A.,   {Kaspi} V.~M.,  eds, 40 Years of Pulsars: Millisecond
  Pulsars, Magnetars and More Vol.~983 of American Institute of Physics
  Conference Series, {The GBT350 Survey of the Northern Galactic Plane for
  Radio Pulsars and Transients}.
pp 613--615

\bibitem[\protect\citeauthoryear{Horne 
\& Baliunas}{1986}]{hb86}
Horne, J.~H., Baliunas, S.~L.\ 1986, ApJ, 302, 757 

\bibitem[Keane et al.(2011)]{kkl+11} Keane, E.~F., Kramer, M., 
Lyne, A.~G., Stappers, B.~W., \& McLaughlin, M.~A.\ 2011, MNRAS, 838 


\bibitem[\protect\citeauthoryear{{Keane}, {Ludovici}, {Eatough}, {Kramer},
  {Lyne}, {McLaughlin} \& {Stappers}}{{Keane} et~al.}{2010}]{klk+09}
{Keane} E.~F.,  {Ludovici} D.~A.,  {Eatough} R.~P.,  {Kramer} M.,  {Lyne}
  A.~G.,  {McLaughlin} M.~A.,    {Stappers} B.~W.,  2010, MNRAS, 401, 1057

\bibitem[\protect\citeauthoryear{{Keane} \& {Kramer}}{{Keane} \&
  {Kramer}}{2008}]{kk+08}
{Keane} E.~F.,  {Kramer} M.,  2008, MNRAS, 391, 2009

\bibitem[\protect\citeauthoryear{{Knight}, {Bailes}, {Manchester}, {Ord} \&
  {Jacoby}}{{Knight} et~al.}{2006}]{kbm+06}
{Knight} H.~S.,  {Bailes} M.,  {Manchester} R.~N.,  {Ord} S.~M.,    {Jacoby}
  B.~A.,  2006, Astrophys.\ J.

\bibitem[\protect\citeauthoryear{{Kramer}, {Lyne}, {O'Brien}, {Jordan} \&
  {Lorimer}}{{Kramer} et~al.}{2006}]{klo+06}
{Kramer} M.,  {Lyne} A.~G.,  {O'Brien} J.~T.,  {Jordan} C.~A.,    {Lorimer}
  D.~R.,  2006, Science, 312, 549

\bibitem[\protect\citeauthoryear{{Li}}{{Li}}{2006}]{li06}
{Li} X.-D.,  2006, ApJL, 646, L139

\bibitem[\protect\citeauthoryear{Lorimer \& Kramer}{Lorimer \&
  Kramer}{2005}]{lk05}
Lorimer D.~R.,  Kramer M.,  2005, {Handbook of Pulsar Astronomy}.
Cambridge University Press

\bibitem[\protect\citeauthoryear{{Luo} \& {Melrose}}{{Luo} \&
  {Melrose}}{2007}]{lm07}
{Luo} Q.,  {Melrose} D.,  2007, MNRAS, 378, 1481

\bibitem[\protect\citeauthoryear{{Lyne}, {Hobbs}, {Kramer}, {Stairs} \&
  {Stappers}}{{Lyne} et~al.}{2010}]{lhk+10}
{Lyne} A.,  {Hobbs} G.,  {Kramer} M.,  {Stairs} I.,    {Stappers} B.,  2010,
  Science, 329, 408

\bibitem[\protect\citeauthoryear{{Lyne}, {McLaughlin}, {Keane}, {Kramer},
  {Espinoza}, {Stappers}, {Palliyaguru} \& {Miller}}{{Lyne}
  et~al.}{2009}]{lmk09}
{Lyne} A.~G.,  {McLaughlin} M.~A.,  {Keane} E.~F.,  {Kramer} M.,  {Espinoza}
  C.~M.,  {Stappers} B.~W.,  {Palliyaguru} N.~T.,    {Miller} J.,  2009, MNRAS,
  400, 1439

\bibitem[McDermott et al.(1988)]{mh88} McDermott, P.~N., van 
Horn, H.~M., \& Hansen, C.~J.\ 1988, ApJ, 325, 725 

\bibitem[\protect\citeauthoryear{McLaughlin, Lyne, Lorimer, Kramer, Faulkner,
  Manchester, Cordes, Possenti, Camilo, Hobbs, Stairs, D'Amico \&
  O'Brien}{McLaughlin et~al.}{2006}]{mll+06}
McLaughlin M.~A.,  Lyne A.~G.,  Lorimer D.~R.,  Kramer M.,  Faulkner A.~J.,
  Manchester R.~N.,  Cordes J.~M.,  Possenti A.,  Camilo F.,  Hobbs G.,  Stairs
  I.~H.,  D'Amico N.,    O'Brien J.~T.,  2006, Nature, 439, 817

\bibitem[\protect\citeauthoryear{{McLaughlin}, {Lyne}, {Keane}, {Kramer},
  {Miller}, {Lorimer}, {Manchester}, {Camilo} \& {Stairs}}{{McLaughlin}
  et~al.}{2009}]{mlk+09}
{McLaughlin} M.~A.,  {Lyne} A.~G.,  {Keane} E.~F.,  {Kramer} M.,  {Miller}
  J.~J.,  {Lorimer} D.~R.,  {Manchester} R.~N.,  {Camilo} F.,    {Stairs}
  I.~H.,  2009, MNRAS, 400, 1431

\bibitem[\protect\citeauthoryear{{Melikidze} \& {Gil}}{{Melikidze} \&
  {Gil}}{2006}]{mg+06}
{Melikidze} G.,  {Gil} J.,  2006, Chinese Journal of Astronomy and Astrophysics
  Suppl. 2, 6, 81

\bibitem[\protect\citeauthoryear{{Miller}, {McLaughlin} \& {Keane}}{{Miller}
  et~al.}{2010}]{mmk+10}
{Miller} J.~J.,  {McLaughlin} M.~A.,    {Keane} E.~F.,  2010, Mon.\ Not.\ R.\
  Astron.\ Soc.

\bibitem[\protect\citeauthoryear{Press, Flannery, Teukolsky \&
  Vetterling}{Press et~al.}{1986}]{pftv86}
Press W.~H.,  Flannery B.~P.,  Teukolsky S.~A.,    Vetterling W.~T.,  1986,
  Numerical Recipes: {T}he Art of Scientific Computing.
Cambridge University Press, Cambridge

\bibitem[\protect\citeauthoryear{{Redman} \& {Rankin}}{{Redman} \&
  {Rankin}}{2009}]{rr09}
{Redman} S.~L.,  {Rankin} J.~M.,  2009, MNRAS, 395, 1529

\bibitem[Reisenegger 
\& Goldreich(1992)]{rg92} Reisenegger, A., \& Goldreich, P.\ 1992, ApJ, 395, 240 


\bibitem[Rosen et al.(2011)]{rr11}
Rosen, R., McLaughlin, M.~A., \& Thompson, S.~E.\ 2011, ApJL, 728, L19

\bibitem[\protect\citeauthoryear{{Scargle}}{{Scargle}}{1982}]{s+82}
{Scargle} J.~D.,  1982, ApJ, 263, 835

\bibitem[\protect\citeauthoryear{Stephens}{Stephens}{1970}]{stephens70}
Stephens M.~A.,  1970, Journal of the Royal Statistical Society, 32, 192

\bibitem[\protect\citeauthoryear{{Wang}, {Manchester} \& {Johnston}}{{Wang}
  et~al.}{2007}]{wmj+06}
{Wang} N.,  {Manchester} R.~N.,    {Johnston} S.,  2007, MNRAS, 377, 1383

\bibitem[\protect\citeauthoryear{{Weltevrede}, {Stappers}, {Rankin} \&
  {Wright}}{{Weltevrede} et~al.}{2006}]{wsrw06}
{Weltevrede} P.,  {Stappers} B.~W.,  {Rankin} J.~M.,    {Wright} G.~A.~E.,
  2006, ApJL, 645, L149

\bibitem[\protect\citeauthoryear{{Woods}, {Kouveliotou}, {Gavriil}, {Kaspi},
  {Roberts}, {Ibrahim}, {Markwardt}, {Swank} \& {Finger}}{{Woods}
  et~al.}{2005}]{wkg+05}
{Woods} P.~M.,  {Kouveliotou} C.,  {Gavriil} F.~P.,  {Kaspi} V.~M.,  {Roberts}
  M.~S.~E.,  {Ibrahim} A.,  {Markwardt} C.~B.,  {Swank} J.~H.,    {Finger}
  M.~H.,  2005, ApJ, 629, 985

\bibitem[\protect\citeauthoryear{{Zhang}, {Gil} \& {Dyks}}{{Zhang}
  et~al.}{2007}]{zgd07}
{Zhang} B.,  {Gil} J.,    {Dyks} J.,  2007, MNRAS, 374, 1103

\end{thebibliography}
\bibliographystyle{mn2e}

\end{document}